\documentclass[aps,pre]{revtex4}
\usepackage{color}

\usepackage{graphicx}
\usepackage{pstricks}

\begin{document}

\title{Computational Study of a Multistep Height Model}
\author{Matthew Drake}
\email[]{matt.drake88@gmail.com}
\affiliation{Department of Physics, University of Massachusetts,
Amherst, MA 01003, USA}
\author{Jonathan Machta}
\email[]{machta@physics.umass.edu}
\affiliation{Department of Physics, University of Massachusetts,
Amherst, MA 01003, USA}
\author{Youjin Deng}
\email[]{yjdeng@ustc.edu.cn}
\affiliation{Hefei National Laboratory for Physical Sciences at Microscale   and Department of Modern Physics,
   University  of Science and Technology of China,  Hefei, Anhui 230026, China
 and Department of Physics, University of Massachusetts, Amherst,
   MA 01003, USA}
\author{Douglas Abraham}
\email[]{d.abraham1@physics.ox.ac.uk}
\affiliation{Rudolph Peierls Centre of Theoretical Physics, University of Oxford,
Oxford OX13NP, UK}
\author{Charles Newman}
\email[]{newman@cims.nyu.edu}
\affiliation{Courant Institute of Mathematical Sciences, New York University,
New York, NY 10012, USA }

\begin{abstract}

\label{sec:abstract}
An equilibrium random surface  
 multistep height model
proposed in [Abraham and Newman, EPL, 86, 16002 (2009)] is studied using a variant of the worm algorithm. 
In one limit, the model reduces to the two-dimensional Ising model in the height representation. When the Ising model constraint 
of single height steps is relaxed, the critical temperature and critical exponents are continuously varying functions 
of the parameter controlling height steps larger than one. Numerical estimates of the critical exponents can be mapped via a single parameter-- the Coulomb gas coupling-- to the exponents of the O$(n)$ loop model on the honeycomb lattice with $n \leq 1$.
\end{abstract}

\maketitle

\section{Introduction}
\label{sec:intro}
Some time ago, two of the authors introduced a statistical mechanical model of a random surface embedded in a three dimensional space, suspended above a planar substrate with which it interacts~\cite{AbNe88,AbNe91,AbFoNePi95}. A detailed specification permitted the exact mapping of configurations of the three-dimensional surface onto those of the planar Ising model in its Peierls-contour representation. When treated by equilibrium statistical mechanics, this model displayed a phase transition; at low enough temperatures, typical configurations indicate that the surface is bound to the substrate. On raising the temperature sufficiently, the interface unbinds from the substrate and a layer of the bulk phase intercalates between the random surface and the substrate. Although it might be thought, as the authors originally did, that this model affords an example of wetting in 2-d \cite{Ab86}, it turns out that the exact critical indices obtained are quite unlike those normally associated with wetting. Rather, this model is much more accurately considered as an example of either Stransky-Krastanov (SK)~\cite{StKr39}, or Volmer-Weber (VW)~\cite{VoWe26} behavior. 

We now go back sixty years to the seminal work of Burton, Cabrera and Frank~\cite{BuCaFr51}, who were the first to propose that a surface phase transition in a 3-d uniaxial system, say the spin-1/2 Ising model, should display singular behavior like the 2-d Ising model.  They considered symmetry-breaking boundary conditions for which boundary spins are fixed to point {\em up} in the upper half space and {\em down} in the lower half space.  They reasoned that below the 3-d critical temperature the upper half space will be in the up magnetized state and the lower half space in the down magnetized state.   In the mean field approximation, the 2-d lattice of spins at the interface between the oppositely magnetized bulk phases is subjected to a mean field that cancels out.  Hence this layer should be strictly 2-d in zero magnetic field and behave accordingly.  This scenario is known to be incorrect; the BCF transition as originally conceived is actually a roughening transition~\cite{WeGiLe73} with quite different characteristics.  The work in~\cite{AbNe88,AbFoNePi95} affords examples in which the Burton, Cabrera and Frank type of  transition is obtained by exact analysis, but in a different scenario and without making a mean field approximation.

Our first step will be to describe the model as originally formulated~\cite{AbNe88,AbNe91,AbFoNePi95} 
and to review the results obtained for it. We will then indicate why we think the SK or VW scenarios are more appropriate than the wetting one. After that, we will point out a number of limitations that were necessary to obtain exact results and then show how Monte Carlo simulations can be used to get information when these limitations are relaxed and the exact result route is no longer available to us.

In the spirit of Kossel and Stransky (KS)~\cite{Ko27,St28}, configurations of molecules adsorbed on the substrate plane are constructed by regarding each molecule as a unit cube the lower side of which meshes with the underlying simple square lattice of the substrate, denoted $\Lambda \subset Z^2$. Molecular rafts are assembled as close packed arrays of the KS cubes. It is clear that, because it has no interior holes, a raft can be described just as well as a simple closed loop on $\Lambda$. Loops can meet at vertices of  $\Lambda$ but not on edges, as this would imply over-counting. The upper faces of the KS cubes have height 1. To extend this model further, we permit placing rafts on top of rafts, without overhangs. In this way, we encounter loops within loops on $\Lambda$. As an additional restriction, the significance of which will soon become apparent, we do not allow any edge of a raft to lie directly above one of any lower raft, thus excluding multiple height jumps. This model thus manifests stepped towers erected on the plane; it was termed the ``multi-ziggurat''  (MZ)~\cite{AbFoNePi95} model in recognition of its architectural precursor in ancient Sumeria. To complete the definition of the MZ model for equilibrium statistical mechanics, we must specify the configurational energy. The surface of a collection of molecular rafts is composed of two parts, one in contact with the substrate and the other in contact with the surroundings. Representing this collection as the equivalent loop form, labelled  $\Gamma$, the energy $E(\Gamma)$  is given by
\begin{equation}
E(\Gamma) = \tau L(\Gamma) + (\tau-\epsilon) A_1(\Gamma).
\label{eq:energya}
\end{equation}
where $\tau$ is the surface tension, 
$\epsilon$ is the adhesion energy of a molecule to
the substrate, and $L(\Gamma)$ is sum of the 
lengths $\ell(\gamma)$ of the simple closed loops of~$\Gamma$,

\begin{equation}
\label{eq:LGamma}
L(\Gamma) = \sum_{\gamma \in \Gamma} \ell(\gamma)
\end{equation}  
The term $A_1$  is the area of contact of the plaquettes  with the substrate, which is exactly the same as the ``roof" area because of the ziggurat construction. Thus the second term in Eq.\ (\ref{eq:energya}) is the energetic contribution of plaquettes in $\Gamma$  with normal perpendicular to the substrate plane. The remainder of the surface tension contribution is given by the first term. For stability against detachment, we evidently require that $\tau > -\epsilon$.  
We now briefly mention  the results that can be obtained for this model. Notice that when $\tau =\epsilon$  we have recaptured precisely the planar Ising model, so the transition temperature and the free energy singularity are known. This is also the case if  $\epsilon>\tau$; then the first layer is completely covered and subsequent rafts are laid on top of this layer as with  $\tau =\epsilon$. In the remaining region of stability, characterised by $-\tau < \epsilon <\tau$, much less detailed information is available~\cite{AbNe91}.

If we are to regard the MZ model as a version of the VW and SK scenarios, then we should point out three shortcomings of the model as it stands. The first, which we will not discuss further here, is that we do not allow for the energy of mismatch, elastic in origin, between the first raft and the substrate. The second deficiency, the examination of which is the main subject of this paper, is the Ehrlich-Schwoebel (ES)~\cite{EhHu66,Sc69} phenomenon. Simply stated, multiple height steps should be allowed, but they are energetically disfavored. In the next section, we will describe in some detail a model of~\cite{collab} that is simply related 
to the MZ model but permits multiple height steps and incorporates the ES idea.

The third problem is the formation of corrals.  A corral or hole is a region of lesser height surrounded by a region of greater height.  
The model studied in this paper forbids corrals.   Detecting a corral requires non-local information and it is unlikely that a physically reasonable equilibrium model would contain long range interactions that forbid corrals.  On the other hand, dynamical considerations may suppress corrals.  The formation of a corral from a plateau requires the removal of a molecule that is entirely surrounded by neighbors.  Such a molecule will be tightly bound and its evaporation suppressed.  A corral might also be formed by the sequential deposition of a wall that eventually encloses a region.  However, surface diffusion would tend to favor more compact structures and suppress the formation of walls.  Thus, although our model is an equilibrium statistical mechanical model, we believe that the no-corrals 
rule makes it an appropriate model for non-equilibrium surface growth processes.   

From the point of view of equilibrium statistical mechanics, the above height models can be regarded as generalizations of 
the height representation of the 2-d Ising model, which corresponds to Eq.~(\ref{eq:energya}) with $\tau= \epsilon$.
Other generalizations are possible. In this work we consider the generalization of~\cite{collab}
that incorporates the ES idea by allowing  height steps 
greater than one. Height steps greater than 
one are given an extra energy  proportional  to $\epsilon_1$. 
We study this  
multistep height model numerically and find that, along its critical curve, the critical exponents vary continuously with $\epsilon_1$.
Though our model is motivated by surface physics, it appears to be closely related to another class of models in statistical physics, the O($n$) loop models.  In the O($n$) loop models the statistical weight depends on the total loop length $L(\Gamma)$ and also contains a factor $n$ for each simple loop.  Thus, the energy $E (\Gamma)$ becomes
\begin{equation}
 E(\Gamma) = \tau L(\Gamma) +   ( \ln n) C(\Gamma) \; ,
\label{eq:energya1}
\end{equation}
where $C(\Gamma)$ is the number of simple loops in $\Gamma$ and $n$ is referred to as the loop fugacity.
On lattices with vertices of degree 3, the loops are disjoint and on the honeycomb lattice one has the well-known O$(n)$ loop model introduced in~\cite{DoMuNiSc81}. 
A great deal is known about this O$(n)$ loop model when $n \leq 2$. 
For a given $n$, there exist three distinct phases: a disordered phase 
with small loops, a densely-packed phase, and a fully-packed phase. 
The densely-packed and fully-packed phases are both critical--i.e., 
the probability for two points to be on the same loop decays algebraically with distance.
Between the disordered and the densely-packed phase is a critical curve as a function of $n$. 
The singular behavior of the O$(n)$ loop model along this critical curve can be described by a set of critical exponents that are functions of $n$.  These exponents can be obtained from a mapping to Coulomb-gas theory~\cite{Nienhuis84}. 
Surprisingly, we find strong numerical evidence that the  multistep height model studied here can also be described 
by the Coulomb gas theory and that there is a one-to-one mapping for 
universal quantities between $n$ and $\epsilon_1$. 
In the continuum scaling limit when the lattice spacing shrinks to zero,
critical $O(n)$ loop models and hence presumably also critical multistep
height models should be conformally invariant and described by 
Conformal Loop Ensembles --- see, e.g., \cite{ScShWi09} --- with
a parameter value $\kappa$ mapped onto $n$ and $\epsilon_1$.

The plan for this paper is as follows. In Sec.\ \ref{sec:model}, we 
define the multistep
height model. In Sec.\ \ref{sec:quantities}, we 
list quantities of interest for the model. In Sec.\ \ref{sec:algorithm},
we describe the numerical methods that are used in our simulations. 
In Sec.\ \ref{sec:results}, we present our results for the critical 
behavior of the multistep height model. The paper closes  with a 
discussion in Sec.\ \ref{sec:discussion}.

\section{The Multistep Height Model}
\label{sec:model}

\begin{figure}[h]
\includegraphics[width=\columnwidth]{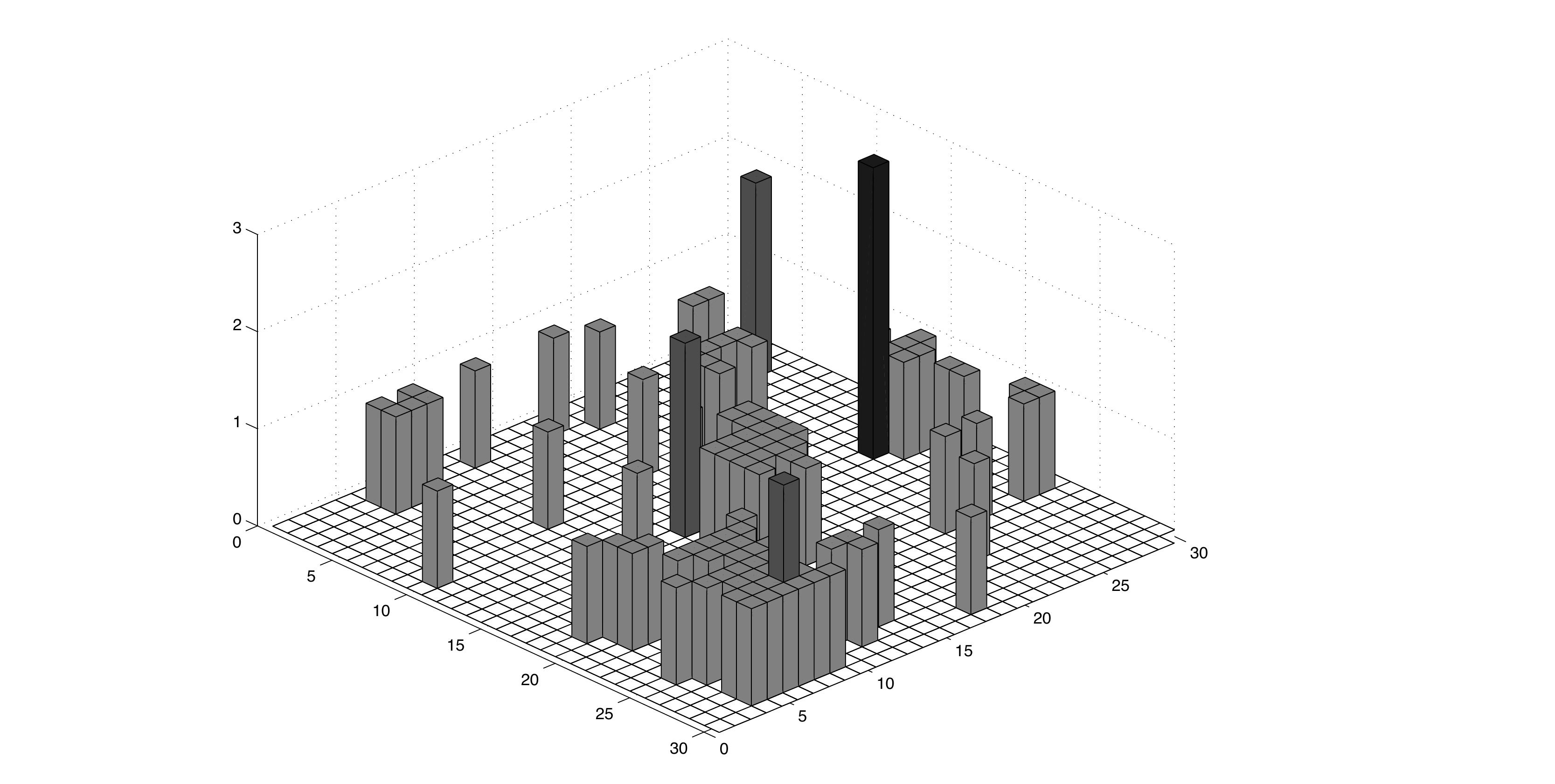}
\label{fig:model}
\caption{A typical configuration of the multistep height model near the critical temperature for $\tau = 2$,  $\epsilon_{1} = 0$ and system size of L=30. The shade of the columns represents their height. }
\end{figure}

The  height models proposed in~\cite{collab} and 
studied here generalize the height representation of the Ising model. 
Consider the two-dimensional Ising model on a square lattice in the spin
representation with uniform plus-spin fixed  boundary conditions.  
There is a one-to-one mapping  from spins on the direct lattice to heights
on the dual lattice.  The height of any dual lattice site is the least
number of Peierls contours that must be crossed on a path from the 
boundary. This definition leads to several important  consequences. 
First, the magnetization of the Ising model is simply the number of 
even height sites minus the number of odd height sites.
Second, there is a constraint that no holes or corrals are allowed in the height representation.  That is, from every dual lattice site, there exists a path to the boundary that is non-increasing in height.  Finally, the height representation of the Ising model will only have height steps of $+1$ or $-1$.

To generalize the model, we break the constraint of single height steps by allowing larger height steps at an extra energy cost, while still disallowing holes or corrals. The probability for a loop configuration is obtained from the energy, $E(\Gamma)$,
\begin{equation}
E(\Gamma) = \tau L(\Gamma) + \epsilon_{1} N(1,1).
\label{eq:energy}
\end{equation}
Here, $\Gamma$ is a collection of simple closed loops defined on the dual
lattice. In the Ising model, these form the Peierls contours, while  in the multistep model they can have a more complicated 
structure. In particular, this model allows Peierls contours to overlap, 
corresponding to  larger height steps. $L(\Gamma)$, previously defined 
in Eq.\ (\ref{eq:LGamma}), is  the sum of the lengths of all of the loops
(sum of all the edge weights), including now the lengths of all of 
the overlapping loops. In the loop representation $\tau$ is the energy 
associated with adding a single edge to $\Gamma$ while $N(1,1)$ is defined
as $L(\Gamma)$ minus the length of all edges with weight one (step size
one). $\epsilon_{1}$ is the energy cost associated with these larger 
height steps. 
$L(\Gamma)$ adds factors of $\tau$ linearly in the size of the height 
step, while $N(1,1)$ adds factors of $\epsilon_{1}$ linearly in the size
of the step but only for height steps larger than one.  Figure 1 shows a typical configuration of the multistep height model near the critical temperature for  $\tau=2$, $\epsilon = 0$, and system size $L=30$.  Note that we have not included the term $(\tau-\epsilon) A_1(\Gamma)$ in Eq.\ (\ref{eq:energya}) in the multistep height.  It would be interesting to consider its effect in future studies.

Equation (\ref{eq:energy}) completely specifies the loop representation of the model.  The height representation is uniquely obtained from the loop representation.  The height associated with a lattice site is obtained using the rule that the height change across  a dual edge is equal to the weight of the bond.  The direction of the height change is determined by the no-corrals rule.   Figure \ref{fig:worm} shows a possible edge configuration and the associated heights.  In this configuration  $L(\Gamma)=62$, $N(1,1)=6$ and $E(\Gamma)=62\tau + 6\epsilon_1$.

In this paper we simulate 
multistep height models for several values of $\epsilon_1$ in the range $-1  \leq \epsilon_{1} \leq \infty$ and $\tau=2$.  Equation (\ref{eq:energy}) with $\tau=2$ and $\epsilon_{1} = \infty$ corresponds to the height representation of the Ising model with interaction energy $J=1$. 

\section{Measured Quantities}
\label{sec:quantities}

Order parameters for height models \cite{orderparam,orderparam2,orderparam3} can be constructed in analogy to the magnetization for spin models. In the height representation of the Ising model,  plus spins correspond to even heights while minus spins correspond to odd heights. One can define  magnetization-like quantities, $M_n$ for positive integers $n$ as is done in \cite{orderparam3},
\begin{equation}
M_{n} = \frac{1}{N} \sum_{j} \exp(\frac{2i\pi h(j)}{n+1}),
\label{eq:mag}
\end{equation}
where $h(j)$ is the height at site $j$ and $N$ is the number of sites. This family of order parameters is motivated by the idea that in the rough (high temperature) phase, any height is nearly equally likely to occur so $M_{n} \rightarrow 0$. In the smooth (low temperature) phase, a single height will dominate and $M_{n} \rightarrow 1$ as $T \rightarrow 0$.  
The magnetization $m$ for the Ising model in the spin representation is equal to $M_1$.  In the critical region of the multistep model, we find that only $M_{1}$ is useful for the small systems studied here. Heights significantly greater than two do not occur often, therefore $M_{n}$ will have strong finite-size corrections for $n>2$.  Henceforth we use $m$ to represent either the magnetization for the spin representation of the Ising model or $M_1$ for height representations.

We measured the following quantities obtained from the height field $h(j)$:
\begin{description}
  \item[Binder cumulant]
  The Binder cumulant $U$ is defined as
 \begin{equation}
U = 1 - \frac{\langle m^{4} \rangle}{3 \langle m^{2} \rangle ^{2}}.
\label{eq:binder}
\end{equation}
Crossings of the Binder cumulants as a function of temperature for various system sizes are used to locate the critical point.
\item[Order parameter] 

From the finite-size scaling of the order parameter $\langle m \rangle$ at the critical temperature we obtain the exponent ratio $\beta/\nu$ from $\langle m \rangle \sim L^{-\beta/\nu}$ where $L$ is the system length.

\item[Susceptibility]  The susceptibility is defined as
\begin{equation}
\chi = \beta N  \langle m^{2} - \langle m \rangle^2 \rangle.
\label{eq:susceptibility}
\end{equation}
From the finite-size scaling of the susceptibility $\chi$ at the critical temperature we obtain the exponent ratio $\gamma/\nu$ from $\chi \sim L^{\gamma/\nu}$.  At the critical temperature for fixed boundary conditions the finite-size scaling relation for $\chi$ is  better behaved if the mean is set to zero, the infinite system critical value. Our finite-size scaling fits for $\gamma/\nu$ are made with the subtraction term $\langle m \rangle^2$ omitted. 
  \item[Specific heat]  The specific heat $c$ is defined as
\begin{equation}
c = \frac{\beta^{2}}{N} (\langle E^{2} \rangle - \langle E \rangle ^{2}).
\label{eq:specificheat}
\end{equation}
From the finite-size scaling of the specific heat $c$ at the critical temperature we obtain the exponent ratio $\alpha/\nu$ from $c \sim L^{\alpha/\nu}$.
\item[Bare substrate areal fraction]  The areal fraction of the bare substrate $\phi$ is defined as
\begin{equation}
\phi =  \frac{1}{N} \sum_{j=1}^{N} \delta_{h(j),0}.
\label{phi}
\end{equation}
For $T \ll T_{c}$ there are almost no ad-atoms and $\phi \approx 1$ while for $T \gg T_{c}$, $\phi \approx 0$.  The bare substrate is also known as the ``gasket" in \cite{ScShWi09} where its (mean) fractal dimension at criticality is calculated rigorously 
for conformal loop ensembles and agrees with earlier non-rigorous results of Duplantier~\cite{Du90} for the area of connected regions in critical O$(n)$ loop models.   Following the notation of Ref.\ \cite{LiDeGa11} the fractal dimension of these domains is  $2-\beta'/\nu$ so that the 
finite-size scaling of $\phi$ is given by $\phi \sim L^{-\beta'/\nu}$. 
\item[Height]
We measured the height at the origin $h(0)$ and the average height of the lattice $\bar{h} = (1/N) \sum_i h(i)$. These quantities are expected to diverge logarthmically in the system size.
\end{description}

\section{Numerical Methods}
\label{sec:algorithm}

We developed two Monte Carlo algorithms to simulate these height models.  For the Ising case, we simulate the spin representation using the Wolff algorithm adapted to fixed boundary conditions. The spins are then mapped to heights in order to measure properties of the height model. For the  multistep height model, there is no spin representation and we use a variant of the worm algorithm to sample the collection of closed loops $\Gamma$.  The non-locality of the no-corrals rule creates significant computational difficulties in implementing the worm algorithm for the multistep height model. 

\subsection{Wolff Algorithm}
\label{subsec:wolff}

The standard Wolff algorithm \cite{wolfforig} for the Ising model must be modified for fixed boundary conditions.  Our approach is to reject clusters that add a boundary spin to the cluster. Here are the steps of the modified Wolff algorithm:

\begin{enumerate}
\item Choose a random site $i$ to initiate the cluster.
\item Using a breadth-first search, consider all neighbors $j$ of every site in the cluster and propose to add $j$ to the cluster.
\begin{itemize}
\item If the neighbor $j$ has the same spin as the cluster add this site to the cluster with probability $p=1 - e^{-2\beta}$.
\item If the site that has been added to the cluster is part of the fixed boundary, reject the entire cluster.  (The rejection of clusters that connect to the boundary is the only new feature required for fixed boundary conditions. It reduces the efficiency of the algorithm compared to free or periodic boundary conditions.)
\item Continue adding sites to the cluster until either the cluster is rejected because it touches the boundary or the growth of the cluster terminates.
\end{itemize}
\item Flip the cluster with probability 1.
\item Collect statistics.
\end{enumerate}

In order to collect statistics on the height field we must map the spin configuration to a height configuration. We do this using a series of breadth-first searches, starting from the boundary.

\begin{enumerate}
\item All the boundary sites are put in the {\em current queue}  and assigned height zero.  All other sites have no assigned height.
\item Do a breadth first search starting from the current queue.  Sites are removed from the current queue one at a time and all neighbors of the site are tested.  Neighbors with the same spin as the current queue and no assigned height  are added to the current queue and given the same height as the current queue.  Neighbors with opposite spin and no assigned height value are assigned a height value one greater than that of the current queue and added to the {\em future queue}. This process is repeated until the current queue is empty.
\item The sites in the future queue are moved to the current queue and the future queue is emptied. 
\item Steps 2 and 3 are performed iteratively until the heights of the entire lattice have been determined.
\end{enumerate}

Although the fixed boundary 
conditions and the spin-to-height mapping slow down the computation, the Wolff algorithm is still very efficient for the height representation of the Ising model with fixed boundary conditions. 
It is significantly faster than  either the single-spin Metropolis algorithm or the worm algorithm described below for the multistep height model.  However, the standard worm algorithm for the Ising model \cite{worm1}  might be more efficient for fixed boundary conditions but was not used for this study.

\subsection{Worm Algorithm}

The worm algorithm was developed as a local update algorithm that, nonetheless, almost entirely eliminates critical slowing down \cite{worm1,worm2,DeGaSo07}. An additional benefit of using the worm algorithm for the multistep height model is the ability to efficiently detect and reject moves that would create corrals. The worm algorithm generates a biased random walk that creates directed edges on the dual lattice and samples the closed loop structures $\Gamma$ with probabilities associated with the energy defined in Eq.\ (\ref{eq:energy}). 

Height models may be represented either by oriented or unoriented loops.  The energy is the same in either case but by using oriented loops it is possible to locally encode the direction of height steps and it is more straightforward to compute heights and detect corrals.
In order to easily satisfy the no corrals rule, the worm algorithm actually samples configurations of 
oriented loops, $\Gamma^\prime$.  We define height changes across an edge relative to the direction that an edge is approached: crossing a right-pointing edge corresponds to an increase in height while crossing a left-pointing edge is a decrease in height. (If $\hat{e}$ is the direction of an edge and $\hat{v}$ the direction the edge is crossed then the direction of the height step is $\hat{e} \times \hat{v}$.) As a result counter-clockwise loops enclose mounds while clockwise loops enclose holes or corrals and clockwise loops are forbidden by the no-corrals rule.  Figure \ref{fig:worm} shows an example of an allowed set of oriented loops and the corresponding heights.

The worm algorithm generates properly weighted, oriented, counter-clockwise  loop configurations as follows:
\begin{figure}[b]
\includegraphics[width=5in]{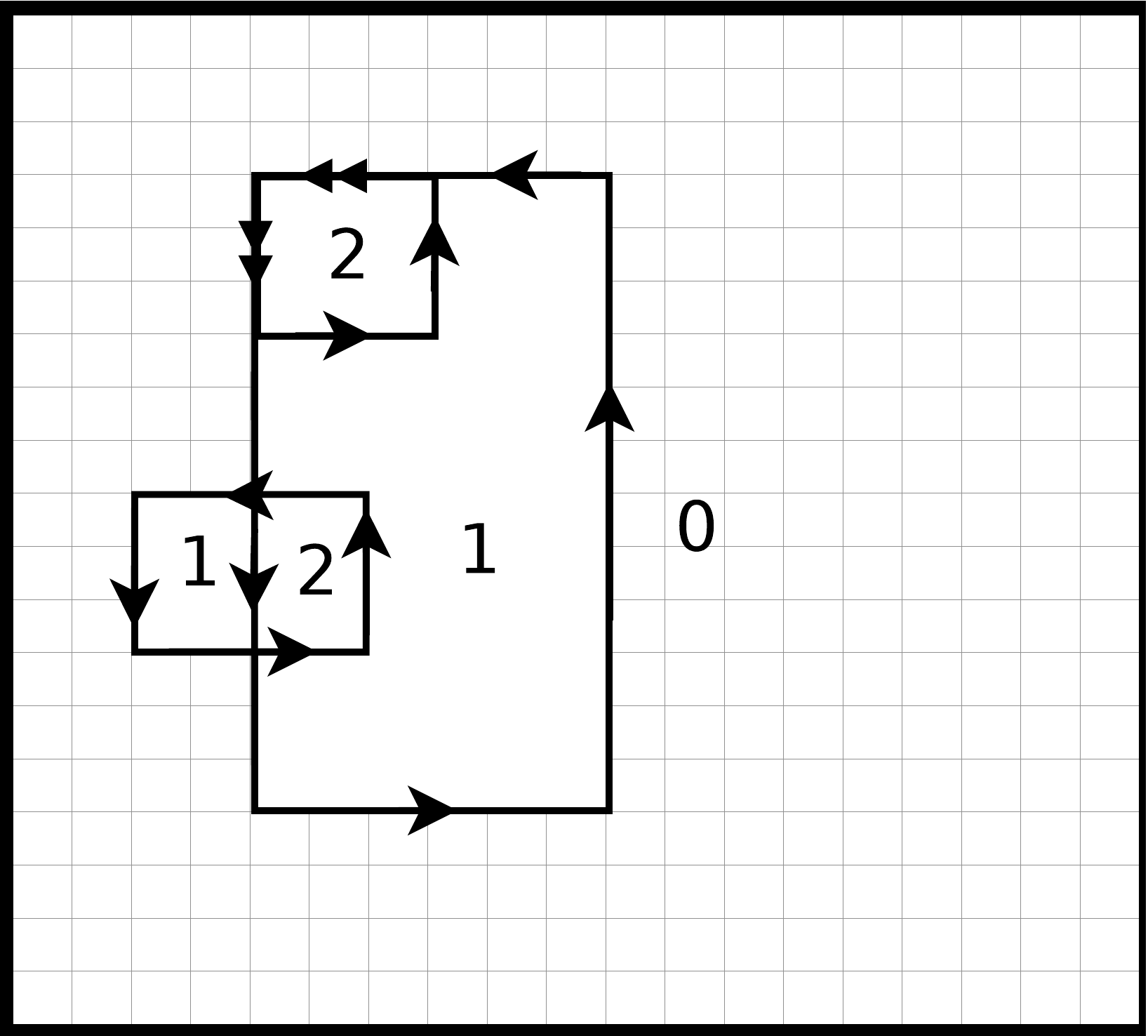}
\caption{A configuration of the multistep height model in the oriented loop represenation.  Each edge corresponds to a height change given by its weight. A double arrow correspond to an edge with weight two, while a single arrow corresponds to an edge with weight one. Crossing an edge with an arrow pointing to the right of the crossing direction is a positive height step. Numbers indicate the heights of the regions.  For this configuration  $L(\Gamma) = 62$ and $N(1,1) = 6$ and $E(\Gamma)=62\tau + 6\epsilon_1$.}
\label{fig:worm}
\end{figure}

\begin{enumerate}
\item Initiate the worm at a random location on the dual lattice. Set the head and tail of the worm at the same position on the dual lattice.
\item Grow the worm. With equal probability, propose to move the head or tail a single step in a random direction along the dual lattice. The movement of the head creates a directed edge pointing from the original position of the head  to its new position. The movement of the tail creates a directed edge pointing from the new  position of the tail to  its original position.  Thus, as the head and tail move, the path connecting the head and tail is created or destroyed.
\item For the proposed move to be accepted, two conditions must be met: the move must be accepted on energetic grounds and a corral must not be created. 
\begin{enumerate}
\item Calculate the change in energy $\Delta E$ associated with the move and the associated Boltzmann factor $e^{-\beta \Delta E}$.  An increase in the edge weight from zero to one is provisionally accepted with probability $e^{-\beta\tau}$ while an increase from a nonzero value to one higher is provisionally accepted with probability $e^{-\beta(\tau+\epsilon_1)}$.  Decreasing the edge weight is always provisionally accepted.  These acceptance probabilities insure that the probability distribution for loop configurations satisfies detailed balance with respect to the set of worm moves.
\item Verify that no corral is created.  As noted previously, this means that no clockwise loop is formed. If the proposed edge joins two existing edges, then a depth-first search that follows the most clockwise edges is carried out. To detect a clockwise loop, we assign every vertex $i$ of the dual lattice an angle $\theta_{i}$. This angle is measured from the point at which the search starts. Right (left) turns yield changes in angle of $-1$ ($+1$).  If the search returns to the initial point and has $\Delta \theta_{i} = -4$, then it is evident that a clockwise loop has been formed and the move is rejected. The search ends by returning to the initial point or when all edges connected to the initial point have been explored.
\end{enumerate}
\item If the edge is provisionally accepted and does not create a corral, the proposed move is accepted and the edge is added to $\Gamma^\prime$. 

\item Repeat steps 2--4 until the head and tail of the worm meet.   It is only when the head and tail meet that the edge set is a  collection of loops and thus a physical configuration.

\item Heights are measured  when the worm closes and $\Gamma^\prime$ is physical. The height of each site is found by scanning through the lattice, row by row, and examining the changes in height by counting the edges that are crossed. The direction of the edge that is crossed determines whether the height change is positive or negative, while the weight of the edge determines the change in height in crossing the edge. All other observables are obtained from the heights.
\end{enumerate}

We find that of these two algorithms, the Wolff algorithm applied to the spin representation followed by a mapping to the height representation is the most efficient way to study the height representation of the Ising model. The worm algorithm and the Wolff algorithm are  comparably  efficient as measured in Monte Carlo sweeps, but the worm algorithm has a substantial computational overhead associated with checking the orientation of loops.  However, the Wolff algorithm is  applicable only in the Ising case where a spin representation is available.  The multistep height model has no known spin representation and thus it is necessary to work either in the height representation or the loop representation.  Both the worm and Wolff algorithms are far more efficient than a single site height update algorithm such as the single-spin Metropolis algorithm or loop algorithms that update loops locally.

For the worm algorithm time can be measured in the number of worms attempted.  A worm attempt consists of choosing a random site and growing the worm from this site until it either closes to form a loop or disappears.  Measured in units of worm attempts, we found that observables approach their equilibrium values exponentially with a time scale that is approximately independent of both the observable and  the value of $\epsilon_1$. This equilibration time scale grows with system size but, for the largest system $L=199$ it is less than 7500.  For all system sizes we discard the first 10$^5$ worm attempts to achieve equilibration before collecting data.  Data is collected every 200 worm attempts.

Errors in the measurement of observables are estimated using the blocking method \cite{NB}. Each simulations consist of 200,000 data collection sweeps ($4\times 10^7$ worm attempts) after the initial equilibratlon. The data is divided into 200 blocks each containing 1,000 measurements and errors are obtained from the standard deviation between blocks. The critical temperature is estimated by extrapolating the crossings of the Binder cumulants to infinite system size as described below. The error in the critical temperature was obtained by considering power law fits of critical observables near this extrapolated critical temperature. The range of temperatures for which the power law fit was a good fit determined the error in the critical temperature. Errors in extrapolated quantities such as the critical exponents are estimated from the range of fit parameters found by varying the critical temperature over its uncertainty range.  Statistical errors in the fits needed to obtain exponents are much smaller, typically 10\% to at most 25\% of the uncertainty resulting from the uncertainty of the critical temperature.

\section{Results}
\label{sec:results}

\subsection{Critical Ising model in the height representation ($\epsilon_{1}=\infty$)}

To understand critical properties of the height representation of the Ising model, the Wolff algorithm with fixed boundary conditions, as described in section \ref{subsec:wolff}, is used along with  finite-size scaling.  We simulated the system at the critical temperature $T_{c}=2/\ln(1+\sqrt{2})$ for 12 sizes ranging from $L = 19$ to $L = 199$ to determine the scaling behavior of the average height per site.   Odd system sizes are used so that the domain has a unique center.

\begin{figure}[h]
\includegraphics[width=4in]{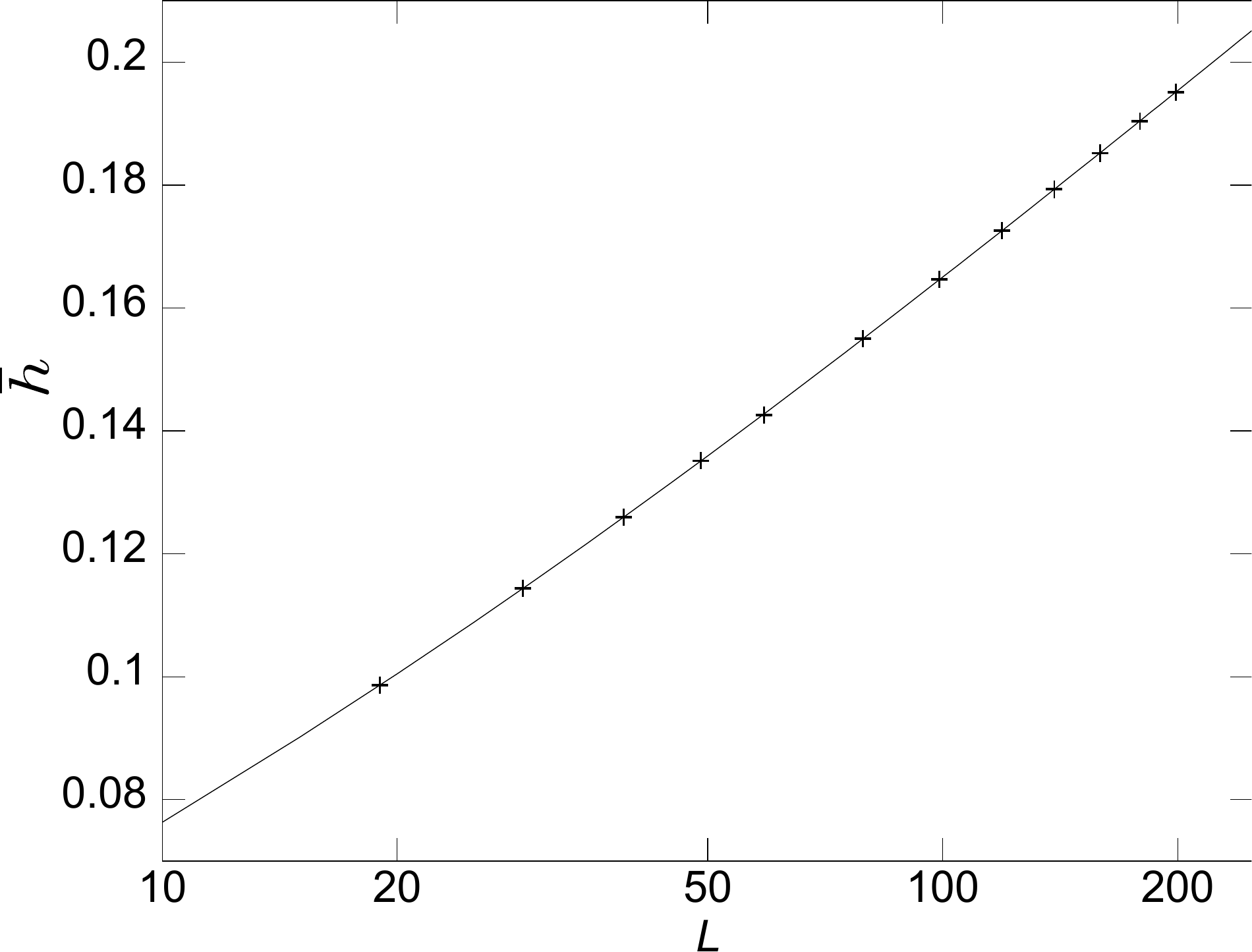}
\caption{The average height $\bar{h}$ versus $L$ for the Ising model.
The solid line is the best fit to the form given in
 Eq.\ (\ref{eq:logfit}) where the prefactor to the logarithm $B = .0467$.}
\label{fig:cgraph}
\end{figure}
Figure \ref{fig:cgraph} shows how the average height per site scales with system size.  The fitting function  is
\begin{equation}
\bar{h} = A + B \ln(L) (1 + \frac{C}{L}).  
\label{eq:logfit}
\end{equation}
The  $1/L$ correction to scaling is suggested by the fixed boundary conditions.
It can be difficult to distinguish between logarithms  from small power laws, but the data clearly shows that this is a logarithmic function. A three parameter
power law fit for $\bar{h}$ yields a Q-value less than .0001, while the 
logarithmic fit of Eq.\ (\ref{eq:logfit}) has a Q value of 0.34.
The prefactor of the logarithm is $B = 0.0467 \pm .0015.$ 
This prefactor is the same within error bars for $\bar{h}$ as it is for $h(0)$.

Using finite-size scaling we also obtain values for 
$\beta/\nu$, $\beta'/\nu$ and $\gamma/\nu$ as shown in Table~\ref{tbl:crit}.  These and other critical  exponents are obtained from a three parameter fit of the form $y = a L^b(1 + c/L)$ where, again the linear correction to scaling is suggested by the fixed boundary conditions and yields reasonable fits.  Although fixed boundary conditions results in larger finite-size corrections than free or periodic boundary conditions we are still able to obtain accurate exponent values.

\subsection{Critical Behavior of the Generalized Height Model ($\epsilon_{1}<\infty$) }
\begin{figure}[h]
\includegraphics[width=5in]{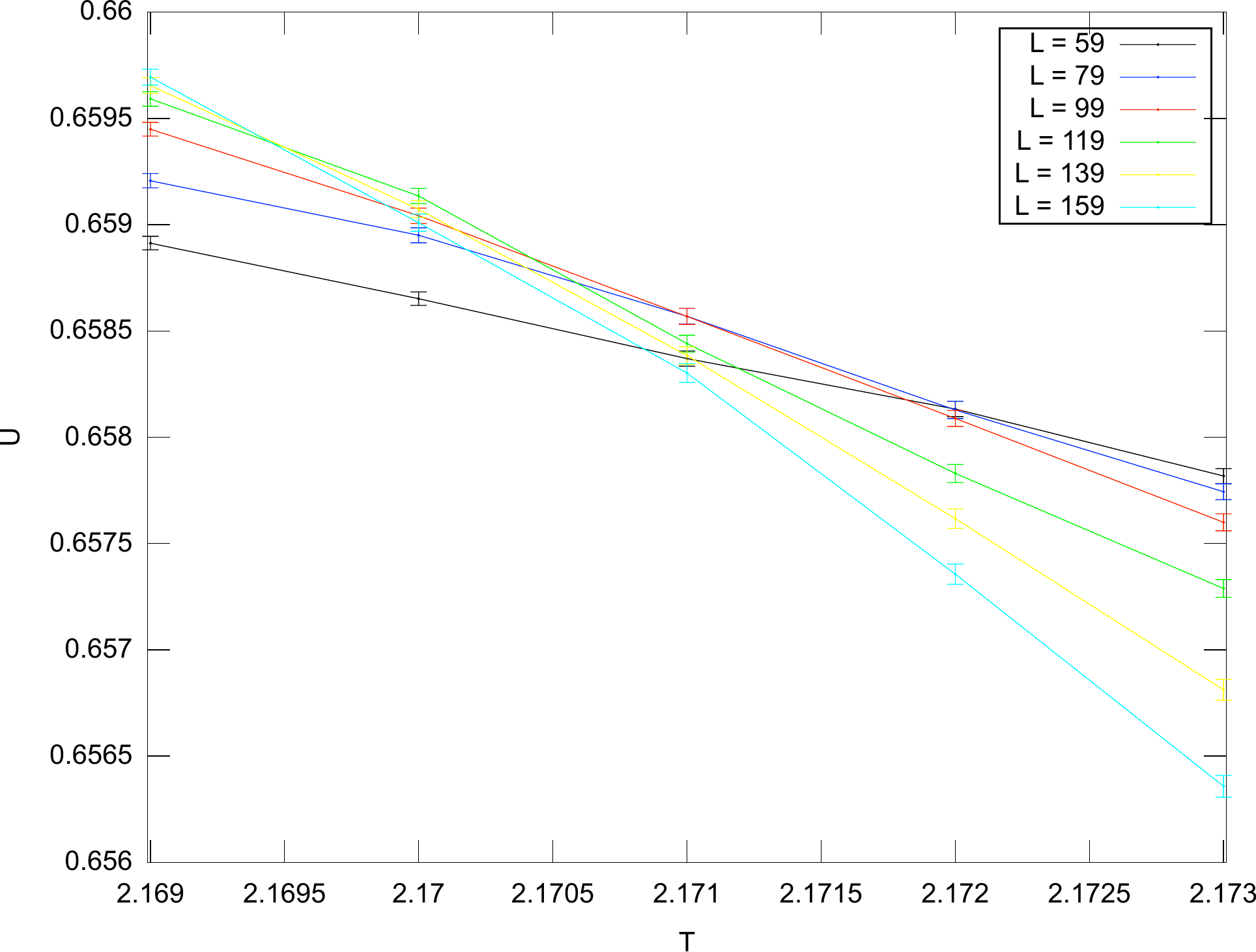}
\caption{(color online)The Binder cumulant $U$ is shown as a function of temperature $T$ for various system sizes and  $\epsilon_{1}=0$.  System sizes increase from top to bottom on the right side of the graph and from bottom to top on the left side.  Finite-size estimates of the critical temperature are obtained from the crossings of the curves for successive system sizes.}
\label{fig:bindergraph}
\end{figure}

\begin{figure}[h]
\includegraphics[width=5in]{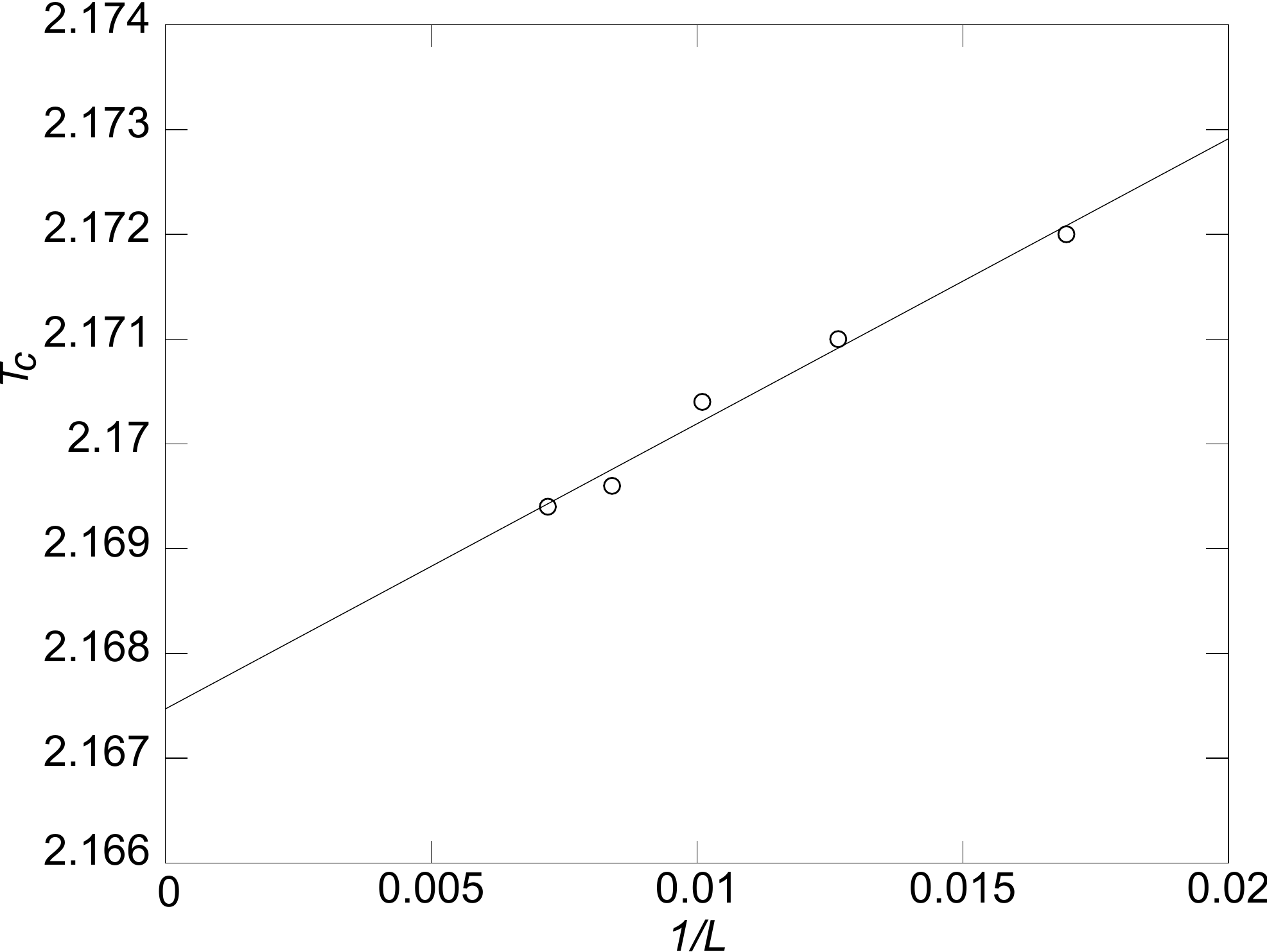}
\caption{The crossings of $U$ are plotted vs. $1/L$ for systems between $L=60$ and $L=160.$ The $L$ value used for each point is the smaller of the system sizes.}
\label{fig:binderextrap}
\end{figure}

The critical temperature of the multistep model depends on $\epsilon_{1}$.
Critical temperatures are obtained from crossings of the Binder cumulant
defined in Eq.\ (\ref{eq:binder}).  Figure \ref{fig:bindergraph} shows the
Binder cumulant for various system sizes.  
Crossings of these curves for successive system sizes gives a finite-size estimate of the critical temperature.  By plotting the crossings 
vs.\ $1/L$ 
and extrapolating to $1/L \rightarrow 0$, we find an estimate of the 
critical temperature in the thermodynamic limit, as seen in 
Fig.\ \ref{fig:binderextrap}. This method is used for various values of
$\epsilon_{1}$. The results are shown in Table~\ref{tbl:crit}.

Using these values for the critical temperature, simulations are carried out at and near the measured critical temperature for 12 system sizes from $L=19$ to $L=199$ to determine the finite-size scaling behavior of each of the quantities of interest. 

\begin{table}[h!]
\begin{tabular}{| c | c | c | c | c | c | c |}
\hline
$\qquad \epsilon_{1} \qquad$ & $T_{c}$ & $\beta/\nu$ & $\gamma/\nu$ & $\alpha/\nu$  & $\beta'/\nu$ & $B$ \\ \hline
$-1$ & 2.0875(10) & .065(3) & 1.868(8) & .33(3)  & .0317(15) & .031(2)\\ \hline
$0$ &  2.1685(10) & .075(2) & 1.85(1) & .26(2)  & .034(1)& .0326(15) \\ \hline
$1$ &  2.2065(10) & .083(2) & 1.83(1) & .21(2)  & .038(2) & .038(2)\\ \hline
$2$ &  2.230(1) & .0925(20) & 1.815(5) & .17(2)  & .042(1)& .040(3) \\ \hline
$\infty$ (simulation) & 2.2686(8) & .127(4) & 1.75(2) & ---  & .053(2)& .0467(15) \\
$\infty$ (exact) & 2.269185\ldots & .125 & 1.75 & 0 & .05208...  & .04594... \\ \hline
\hline
\end{tabular}
\caption{Critical temperature $T_c$, critical exponents $\beta/\nu$, $\gamma/\nu$, $\alpha/\nu$ and  $\beta'/\nu$, and the prefactor of the logarithmic scaling of the average height $B$ for each simulated value of $\epsilon_1$.
The Ising model corresponds to $\epsilon_1=\infty$.  Ising model results were obtained from the Wolff algorithm.  Exact results for the Ising model are presented for comparison.}
\label{tbl:crit}
\end{table}

We find that the multistep model has critical behavior that is dependent on the value of $\epsilon_{1}$. We study the exponents $\beta/\nu$, $\gamma/\nu$, $\alpha/\nu$, and $\beta'/\nu$, as well as the logarithmic prefactor $B$ of the average height per site. Table \ref{tbl:crit} lists the resulting fits for each value of $\epsilon_{1}$.  
The errors in the specific heat measurements are too large to discriminate between a logarithmic scaling law ($\alpha/\nu = 0 (\log)$) and a small power law $\alpha/\nu > 0$. Both fits are plausible and yield similar goodness of fit values.  The values of $\alpha/\nu$ in Table \ref{tbl:crit} assume a power law fit. 

It is possible that the observed continuous variation in the exponents shown 
in Table~\ref{tbl:crit} reflects finite-size corrections and that the number of universality classes is small, perhaps one or two.   However, we believe that the evidence points to the conclusion that the multistep height model describes a one parameter family of universality classes parameterized by $\epsilon_1$.  In two dimensions, there are several parameterized statistical mechanical models that have continuously varying critical exponents. These include the $q$-state random-cluster model 
in terms of parameter $q \in [0,4]$~\cite{FoKa}, 
the O$(n)$ loop model for $n \in [-2,2]$, the Ashkin-Teller model in the associated coupling strength, as
well as the 6- or 8-vertex models. In all of these systems, the critical exponents can be expressed in terms of a single parameter--the 
coupling constant $g$ in the Coulomb-gas model.   Thus, it is plausible that $g$ is also sufficient to describe the critical 
exponents of the multistep height model.

We note that the multistep height model and the O$(n)$ loop model~\cite{DoMuNiSc81,Nienhuis82,DeGaGuBlSo07,LiDeGa11} are similar in several ways.  Both models are formulated in terms of loops and contain the Ising model as a special case.  Both models have statistical weights that depend in the same way on the total loop length, $L(\Gamma)$.   The O$(n)$ loop model is typically defined on a honeycomb lattice and requires Eulerian loops so that loops do not overlap and the statistical weight has a term $n^{C(\Gamma)}$ where $C(\Gamma)$ is the number of simple loops.  When $n=1$ the model reduces to the Ising model.  
In Ref.~\cite{Nienhuis82}, by making use of the fact 
that the critical O$(n)$ loop model is equivalent to the tricritical $q=n^2$ Potts model, 
the critical exponents for the magnetization, the Ising-spin domain, the susceptibility, and 
the specific heat were related by simple formulae to the Coulomb gas coupling. 
In terms of the exponents $\beta/\nu, \beta'/\nu, \gamma/\nu$, 
and $\alpha/\nu$, these formulas read~\cite{Nienhuis82,Nienhuis84,Saleur87,Duplantier89,DeGaGuBlSo07,LiDeGa11}:
\begin{eqnarray}
\beta/\nu &=& \frac{6-g}{g} \; , \label{eq:g1} \\
\beta'/\nu &=& \frac{(g-2)(6-g)}{8g} \; , \label{eq:g2} \\
\gamma/\nu &=& \frac{4g-12}{g} \;  , \label{eq:g4}\\ 
\alpha/\nu &=& 6 - \frac{32}{g} \; , \label{eq:g3}
\end{eqnarray}
with $g \in [4,6]$. Equation~(\ref{eq:g4}) is related to Eq.\ (\ref{eq:g1}) by $\gamma/\nu =2 -2 \beta/\nu$.  
 Later in this section we will discuss the relation between the
constant $B$ in Equation~(\ref{eq:logfit}) and the Coulomb gas parameter $g$. 

We use our measurements of $\beta/\nu, \beta'/\nu, \gamma/\nu ,$ and $\alpha/\nu$ to estimate $g$ for each value of $\epsilon_1$. 
Estimates of $g$ from these exponents agree well with each other and a combined estimate can be obtained from a weighted average of the estimates from each of the four critical exponents.  The weight assigned to exponent $i$ is $1/\delta_{i}^{2}$ where $\delta_i$ is the error in the estimate of the corresponding exponent.
Table \ref{tbl:g} lists the value of $g$, calculated by performing a weighted average of each of the four measured exponents together with the goodness of fit $Q$ for combining the values from each exponent into a single value of $g$.  The goodness of fit is calculated assuming the critical exponents are independent and normally distributed with a standard deviation given by $\delta_i$.  These assumptions are not quantitatively correct so the $Q$ values cannot be interpreted as the probability of finding $\chi^2$ larger than the fit value.  Nonetheless, the fact that the $Q$ values are large suggests that a single  $g$  simultaneously predicts all of the exponents within their error ranges.  The existence of a single $g$ consistent with all the critical exponents supports the idea that the multistep height models parameterized by $\epsilon_1$ are indeed in the universality classes of the Coulomb-gas model parameterized by $g$.  The last column in  Table \ref{tbl:g} lists the value of the loop fugacity $n$ in the O$(n)$ loop model for the value of $g$, which is obtained from the formula \cite{Nienhuis82,Nienhuis84}
\begin{equation}
 n = -2 \cos(\pi g/4).
 \label{eq:nvsg}
\end{equation}  
The relation between $n$ and $\epsilon_1$ for the four finite values of $\epsilon_1$ that were simulated is close to linear and extrapolates to $n=1/2$ for $\epsilon_1=-2$, which is the limit of stability of the multistep height model.  

\begin{table}[h!]
\begin{tabular}{| c | c | c |c|c|}
\hline
$\qquad \epsilon_{1} \qquad$ & $g$(fit) & $Q$ &$g$(conj) &$n$(fit)\\ \hline
$-1$ & 5.629(10) & .759 &5.6261&.574(15) \\ \hline
$0$ &  5.579(8) & .986&5.5759 &.649(11) \\ \hline
$1$ &  5.535(9) & .802&5.5297 &.714(13)\\ \hline
$2$ &  5.488(7) & .731&5.4890 &.783(10)\\ \hline
$\infty$(measured) & 5.324(16)& .977&5.3333&1.013(43) \\ \hline
$\infty$(exact) & 16/3 & --&-- & 1\\
\hline
\end{tabular}
\caption{For each simulated value of $\epsilon_1$, the best fit values of the Coulomb gas coupling  $g$(fit), the goodness of this fit $Q$, the conjectured value of the Coulomb gas coupling $g$(conj) obtained from Eqs.\ (\ref{eq:nvsg}) and (\ref{eq:nvse}), and the  loop fugacity $n$(fit) obtained from $g$(fit) and Eq. (\ref{eq:nvsg}).}
\label{tbl:g}
\end{table}

\begin{figure}[h]
\includegraphics[width=5in]{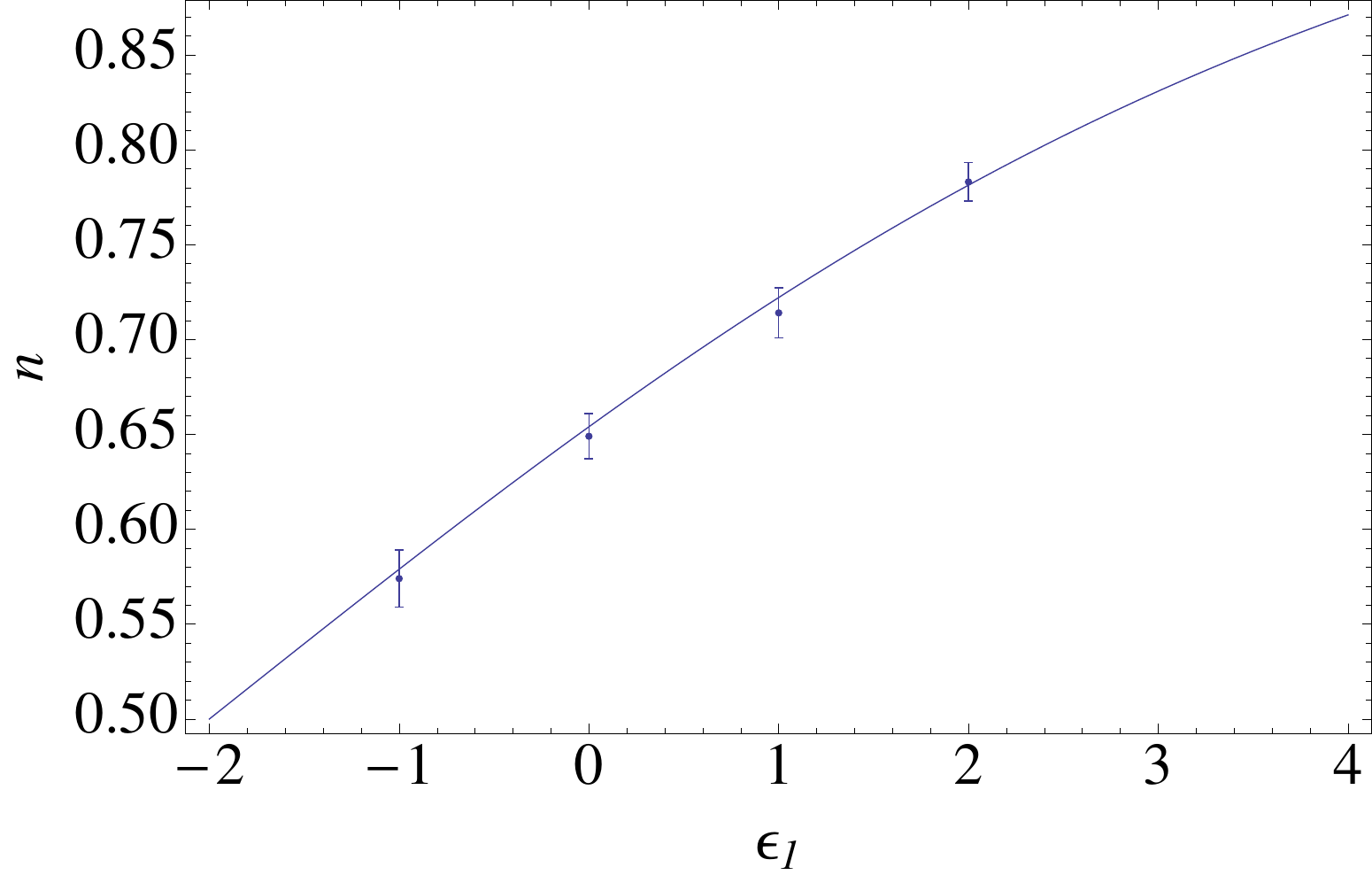}
\caption{The loop fugacity $n$ of the O$(n)$ loop model vs.\ $\epsilon_1$.  The curve is obtained from Eq.\ (\ref{eq:nvse}) and the data is from Table \ref{tbl:g}. }
\label{fig:nvse}
\end{figure}

We find that the simple relation 
\begin{equation}
\label{eq:nvse}
n= \frac{\tanh\left[(\epsilon_1+2)/2 \pi\right]+1}{2}
\end{equation}
is a good fit to the data in Table \ref{tbl:g}.  
The function is a guess based on the requirement that it yields $n=1/2$ for $\epsilon_1=-2$ and approaches
$n=1$ exponentially  as $\epsilon_1 \to \infty$ as might be expected since $\epsilon_1$ exponentially suppresses loop weights greater than one.  Figure \ref{fig:nvse} shows the data in Table \ref{tbl:g} together with the conjectured relation, Eq.\ (\ref{eq:nvse}).   It would be interesting to simulate larger values of $\epsilon_1$ to test this conjecture beyond the linear regime.  
If the conjecture holds, it suggests the possibility of an exact mapping from the multistep height model to a  known 2-d model.   

There are also exact results for the Coulomb gas model for the universal prefactor of the logarithmic growth of the height at the origin.  It is straightforward to show that the same prefactor holds for both the height at the origin and the average height, $\bar{h}$.  The prediction for the logarthmic prefactor $B$ defined in Eq.\ (\ref{eq:logfit}) from \cite{CaZi03}, which has been proved  for conformal loop ensembles~\cite{ScShWi09}, is 
\begin{equation}
\label{eq:B}
B=\frac{(g-4)}{\pi g} \cot(\pi g/4) .
\end{equation}
If we use the values of $g$ given in Table \ref{tbl:g} to compute $B$ using Eq.\ (\ref{eq:B}) and compare with the measured values in Table \ref{tbl:crit}, we find reasonable agreement.   The measured values are larger than the predicted values by about twice the quoted error except for the Ising case where the two values differ by about the quoted error.  Since the measurement of $B$ is likely to have substantial systematic errors that are not included in the $1/L$ correction in Eq.\ (\ref{eq:logfit}), we believe that the results for $B$ are consistent with the Coulomb gas predictions and provide additional evidence that the 
multistep height models are in the universality classes of the Coulomb gas/conformal loop ensemble. 

\section{Discussion}
\label{sec:discussion}

We have analyzed a height model that generalizes the height representation of the Ising model by allowing height steps greater than unity with an energy cost  parameterized by $\epsilon_1$. Our data supports the hypothesis that the critical exponents and prefactor $B$ depend continuously on  $\epsilon_{1}$. Although we believe that the system has continuously varying critical exponents,  it is conceivable that there are only one or two universality classes and that the $\epsilon_{1}$ dependent exponents reflect a finite-size crossover.  The hypothesis of continuously varying exponents is strengthened by the relationship between the measured exponents that allows a  mapping via the Coulomb gas parameter 
$g$ 
to the  O$(n)$ loop model for $n$ in the range $1/2 \leq n \leq 1$.  We 
conjecture an exact relationship between $\epsilon_1$ and $g$ or $n$ that merits further investigation.

In the multistep height model the energy of a height step is a linear function of its magnitude.  It would be interesting to study other energy functions.  For example, the energy of a height step could be quadratic in its magnitude.   We suspect that this would also yield models in the Coulomb gas universality classes. 

A central feature of the model studied here is the no-corrals rule. It would be interesting to study a model with the same energetics but without forbidding corrals. Height models that allow corrals are generally referred to as ``solid-on-solid" models.   If the energy for height steps greater than one is quadratic and corrals are allowed the corresponding discrete Gaussian model is known to be in the $O(2)$ universality class \cite{Nienhuis87}. 
The body-centered solid-on-solid (BCSOS) model restricts height steps to one and has an energy that is a sum of  next-nearest neighbor height differences.  The BCSOS model is exactly solvable and is also in the $O(2)$ universality class \cite{vBeijeren77,Baxter}. 
In the loop representation, allowing corrals corresponds to 
oriented loops with both orientations allowed. 
In the case of oriented loops on the honeycomb
lattice with $\epsilon_1 = \infty$, the loops are disjoint and
the orientation degrees of freedom can be simply integrated out.  This yields a statistical weight of $2$ for each loop, and hence
maps to the $O(n)$ loop model with $n=2$. On this basis, it is tempting to speculate that for $0 < \epsilon_1 < \infty$, these oriented loop models will map to $O(n)$ loop models with loop fugacity in the range
$1< n < 2$.

\acknowledgements
We acknowledge useful discussions with Nikolay Prokofiev.  J. M. and M. D. were supported by NSF DMR-0907235.  
C.N. was supported by NSF OISE-0730136 and NSF DMS-1007524..   
Y. D. was supported by the NSFC under Grant No. 10975127.

%\bibliography{paperbib}

\end{document}